\documentclass[final]{aipproc}
\layoutstyle{6x9}
\def\bge{\begin{equation}}
\def\ene{\end{equation}}
\def\bg{\begin{eqnarray}}
\def\en{\end{eqnarray}}
\usepackage{amsmath,epsfig}
\begin{document}
\title{The quark-meson coupling model and chiral symmetry}
\author{Koichi Saito}{
  address={Department of Physics, Faculty of Science and Technology \\
Tokyo University of Science, Noda, Chiba 278-8510, Japan
}
}

\begin{abstract}
We extend the quark-meson coupling (QMC) model to incorporate chiral symmetry.  
The relationship between the QMC model and chiral perturbation theory is also discussed. 
The nuclear central potential is modified by the effect of internal structure of nucleon. 
\end{abstract}

\maketitle


\section{Introduction}

The quark-meson coupling (QMC) model, first proposed by Pierre Guichon~\citep{qmc}, has been extensively 
developed and applied to various nuclear phenomena with tremendous success~\citep{qmcrev}.  
The QMC model can provide the medium modification of nucleon structure 
based on the quark and gluon degrees of freedom. 

In fact, the evidence for the change of nucleon structure may be observed in polarization transfer measurements 
of the quasi-elastic (${\vec e}$, $e^\prime {\vec p}$) reaction at the Jefferson laboratory and MAMI~\citep{jlab}. 
It also seems vital to consider the internal structure change of nucleon to understand the nuclear EMC effect~\citep{qmcrev,emc}.  

On the other hand, a major breakthrough occurred in the problem of nucleon-nucleon ($N$-$N$) force by introducing the concept of effective 
field theory (EFT) or chiral perturbation theory (ChPT)~\citep{mach}. 
It gives a low-energy scenario that pions and nucleons 
(and possibly deltas) interact via a force
governed by spontaneously broken, approximate chiral symmetry. 

It is thus important to incorporate chiral symmetry into the QMC model, and it may be achieved by using  
the volume-coupling version of the cloudy bag model (CBM)~\citep{cbm,tony}. 
Here, we propose a QMC model with chiral symmetry (we call it 
the chiral QMC (CQMC) model~\citep{cqmc}), and discuss the relationship between the CQMC and ChPT. 

\section{Cloudy bag model with gluon}

Using a unitary transformation, $S = \exp[ i{\vec \tau} \cdot {\vec \phi} \gamma_5 /2 f_\pi]$, 
the chiral bag model, which is given in terms of the MIT bag with the surface interaction 
between the confined quarks and pions~\citep{jaffe}, is transformed to a more practical form, that is the 
volume-coupling version of CBM~\citep{tony}: 
\bge  
{\cal L}_{CBM} =  \left[ {\bar \psi} \left\{ i \ \slash\!\!\!\!\!\!{\cal D} + \frac{1}{2f_\pi} \gamma_\mu \gamma_5 
{\vec \tau} \cdot (D^\mu {\vec \phi}) \right\} \psi -B \right] \theta_V - \frac{1}{2}{\bar \psi} \psi \delta_S 
+ \frac{1}{2} ( D_\mu {\vec \phi} \,)^2 + {\cal L}_{\chi B} , \label{chiralv-lag}
\ene
with $\psi$ the quark field, ${\vec \phi}$ the pion field, 
$\phi = ({\vec \phi}\cdot {\vec \phi})^{1/2}$, $f_\pi$ ($=93$ MeV) the pion decay constant, 
$B$ the bag constant, $\theta_V$ the step function for the bag and $\delta_S$ the surface $\delta$-function.  
Here, ${\cal D}^\mu$ is the covariant derivative, and 
$D_\mu {\vec \phi} = (\partial_\mu \phi) {\hat \phi} + f_\pi \sin(\phi /f_\pi)\partial_\mu {\hat \phi}$. 
The last term, ${\cal L}_{\chi B}$, describes the symmetry breaking terms including the $u, d$ quark mass, $m_0$, and 
the pion mass, $m_\pi (=138$ MeV).

Linearizing the pion field and 
introducing the gluon field, ${\vec A}^\mu$, the Lagrangian density reads 
${\cal L}_{CBM} = {\cal L}_{BAG} + {\cal L}_{\pi} + {\cal L}_{g} + {\cal L}_{int}$, 
where ${\cal L}_{BAG}$ is the usual, bag Lagrangian density, and the interaction is 
\begin{equation}
{\cal L}_{int} = {\bar \psi} \left[ i \frac{m_0}{f_\pi}\gamma_5 {\vec \tau} \cdot {\vec \phi} 
+ \frac{1}{2f_\pi} \gamma_\mu \gamma_5 {\vec \tau} \cdot (\partial^\mu {\vec \phi}) 
+ \frac{g}{2} \gamma_\mu {\vec \lambda} \cdot {\vec A}^\mu  \right] \psi \, \theta_V  , \label{int-lag}
\end{equation}
with ${\vec \lambda}$ the SU(3) color generators and $g$ the quark-gluon coupling constant. 
The free pion field and the gluon are, respectively, described by ${\cal L}_{\pi}$ and ${\cal L}_{g}$.  

The one-gluon exchange (OGE) contribution to the hadron mass is split into the color electric ($\Delta E_G^{el}$) and magnetic 
($\Delta E_G^{mg}$) parts, and the latter is important~\citep{mit}.  This magnetic force differently contributes 
to the $\Lambda$ and $\Sigma$ hyperons, because the $u$ and $d$ quarks in $\Lambda$ ($\Sigma$) are in the 
spin-singlet (triplet) state~\citep{hyp}. 
In fact, $\Delta E_G^{mg}$ lowers both masses of $\Lambda$ and $\Sigma$, but the mass reduction for  
$\Lambda$ is much larger than for $\Sigma$~\citep{hyp2}. 

The second-order energy correction caused by the pion-quark interaction 
in Eq.(\ref{int-lag}) is calculated by the Hubbard's prescription.  In Refs.\citep{cqmc,hyp2}, the Hartree-Fock (HF) contribution to 
the hadron mass is considered. 

In the actual calculation, we fix the $u$, $d$ quark mass, $m_0 = 5$ MeV. Furthermore, we assume that 
the usual $z$ parameters for the $N$ and $\Delta$ are the same, namely $z_0 = z_N = z_\Delta$. 
Then, the bag constant, $B$, and $z_0$ are determined so as to fit 
the free nucleon mass, $M_N (= 939$ MeV), with its bag radius, $R_N = 0.8$ fm. 
The remaining parameter, $g$, is chosen so as to generate the correct mass difference between $M_N$ and the $\Delta$ mass, 
$M_\Delta (=1232$ MeV).  Lastly, we use the strange quark mass, $m_s$, to fit the $\Omega$ mass ($M_\Omega = 1672$ MeV). 
We then study two cases where the pion-cloud effect is included 
(case $1$) or not included (case $2$). The bag parameters for the case $1$ or $2$ are given in Ref.\citep{hyp2}.  

The N-$\Delta$ mass difference (about $300$ MeV) is mainly reproduced by  
the OGE contribution (about $240$ MeV). In contrast, the pion-cloud contribution is 
about $60$ MeV, which is near the upper limit allowed from lattice QCD constraints~\citep{young}.   
We also calculate the mass of hyperon ($\Lambda$, $\Sigma$, or $\Xi$).  In such calculation, we take a different 
$z$ parameter for each hyperon and fit the calculated mass to the observed value in free space~\citep{hyp2}.  

\section{QMC and chiral symmetry}

To describe a nuclear matter, we need the intermediate attractive and short-range repulsive nuclear forces. 
As in the usual QMC model~\citep{qmcrev}, it may be achieved by introducing the $\sigma$ and $\omega$ mesons. 
The present $\sigma$ meson represents, in some way, the exchange of two pions in 
the iso-scalar $N$-$N$ interaction.  

Assuming mean-field approximation (MFA), let us add the Lagrangian density 
\begin{equation}
{\cal L}_{\sigma \omega} = {\bar \psi} \left[ g_\sigma^q {\bar \sigma} - g_\omega^q \gamma_0 {\bar \omega} \right] \psi \, \theta_V  
+ \frac{1}{2} (\partial_\mu {\bar \sigma})^2 - \frac{1}{2} m_\sigma^2 {\bar \sigma}^2 - \frac{1}{2} (\partial_\mu {\bar \omega})^2  
+ \frac{1}{2} m_\omega^2 {\bar \omega}^2 ,  \label{qmc-lag2}
\end{equation}
to ${\cal L}_{CBM}$.  Here, $g_\sigma^q (g_\omega^q)$ is the $\sigma (\omega)$-quark coupling constant, 
$m_\sigma (m_\omega)$ is the $\sigma (\omega)$ meson mass, and ${\bar \sigma} ({\bar \omega})$ is 
the mean-field value of the $\sigma$ ($\omega$) meson in matter.  
The Lagrangian density at the hadron level is then given by~\citep{cqmc}
\bge
{\cal L}_{CQMC} = {\bar N} \left[ i \slash\!\!\!\!\! \partial - M_N^\star - g_\omega \gamma_0 {\bar \omega}  \right] N 
+ \frac{1}{2} (\partial_\mu {\bar \sigma})^2 - \frac{1}{2} m_\sigma^2 {\bar \sigma}^2 - \frac{1}{2} (\partial_\mu {\bar \omega})^2  
+ \frac{1}{2} m_\omega^2 {\bar \omega}^2 , \label{chiral-3}
\ene
with $N$ the nucleon field and $g_\omega = 3 g_\omega^q$.  Note that the classical pion field vanishes in the nuclear ground state. 
The effective nucleon mass, $M_N^{\star}$, is calculated by using the CBM, and it is given as a function of ${\bar \sigma}$: 
\bge
M_N^{\star}({\bar \sigma}) = M_N - g_\sigma({\bar \sigma}) {\bar \sigma} , \ \ \ \ \ \ \ \ 
g_\sigma({\bar \sigma}) = g_\sigma \left[ 1-\frac{a}{2}(g_\sigma {\bar \sigma}) \right] , 
\label{M-star}
\ene
where $g_\sigma = g_\sigma({\bar \sigma} = 0)$ and $a$ is the scalar polarizability. 

In a uniformly distributed nuclear matter, the derivative term of the meson field vanishes.  
Thus, the meson field is no longer dynamical but just auxiliary.  
The classical meson field is thus replaced by the nucleon density, $N^\dagger N$, or the nucleon scalar density, ${\bar N}N$. 
Using Eq.(\ref{M-star}), we find 
\bge
g_\omega {\bar \omega} = \frac{C_v^2}{f_\pi^2} (N^\dagger N) , \ \ \ \ \ \ \ \ 
g_\sigma {\bar \sigma} = \frac{C_s^2}{f_\pi^2} \left[ \frac{({\bar N} N)}{1+ \frac{C_s^2}{f_\pi^2} a ({\bar N} N)} \right] , 
\label{replace}
\ene
where $C_v^2/f_\pi^2 = g_\omega^2/m_\omega^2$ and $C_s^2/f_\pi^2 = g_\sigma^2/m_\sigma^2$.  Then, the Lagrangian density reads 
\bg
{\cal L}_{CQMC} &=& {\bar N} \left[ i \slash\!\!\!\!\! \partial - M_N \right] N + {\cal L}_{4N} + {\cal L}_{hot} ,  \\
{\cal L}_{4N} &=& \frac{C_s^2}{2f_\pi^2} ({\bar N} N)^2 - \frac{C_v^2}{2f_\pi^2} (N^\dagger N)^2 ,  \\ 
{\cal L}_{hot}  &=& - \frac{C_s^4}{2f_\pi^4} a ({\bar N} N)^3 + \frac{C_s^6}{2f_\pi^6} a^2 ({\bar N} N)^4 + \cdots . \label{chiral-N}
\en
This is equivalent to the chiral effective Lagrangian in MFA~\citep{weinberg}. The last term, ${\cal L}_{hot}$, 
gives higher-order terms, which describe the effect of many-body forces as studied in Ref.\citep{skyrme}, and it 
is related to the internal structure of nucleon, namely the scalar polarizability. 
Note that $(C_s^2/f_\pi^2) a ({\bar N} N)$ in Eq.(\ref{replace}) is about 0.3 at 
normal nuclear density, $\rho_0 (= 0.15$ fm$^{-3})$. 

In the numerical calculation for symmetric nuclear matter,  
we take $m_\sigma = 550$ MeV and $m_\omega = 783$ MeV, and 
the $\sigma$-N and $\omega$-N coupling constants, $g_\sigma$ and $g_\omega$, 
are determined so as to fit 
the nuclear saturation condition ($E_{tot} - M_N = -15.7$ MeV) at $\rho_0$~\citep{cqmc,hyp2}. 

\begin{center}
\begin{figure}[t]
\epsfig{file=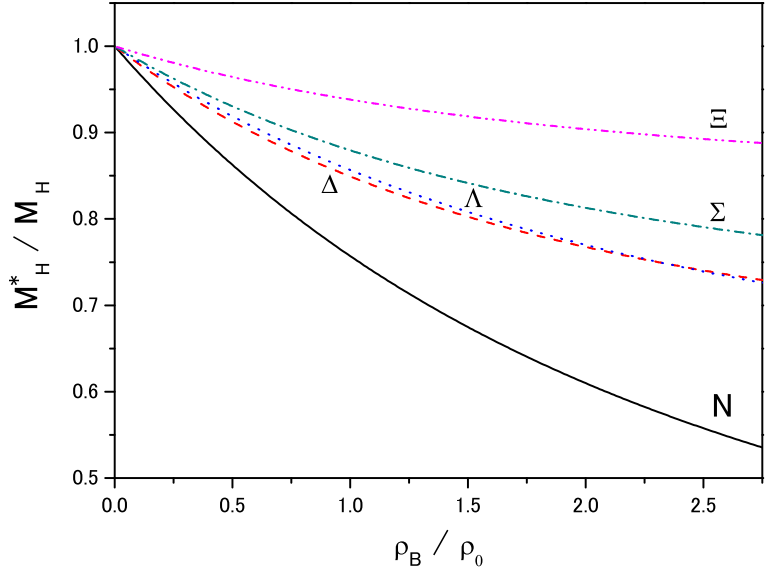,width=12cm}
\caption{Ratio of the baryon mass in matter to the free one (for the case 1). }
\label{fig:rmas}
\end{figure}
\end{center}

In Fig.~\ref{fig:rmas}, we present the baryon mass at the nuclear density, $\rho_B$.  
\noindent
The scaling law for hadron masses in the simple QMC~\citep{scale} is apparently violated by the hyperfine interaction. 
In fact, the $\Sigma$ and $\Lambda$ masses are split up in matter, and 
the $N$-$\Delta$ mass difference is very enhanced by the color magnetic 
interaction.  The pion-cloud effect is relatively minor in the baryon spectra.
We have also discussed the (mean-field) potential for the hyperon in a nuclear matter~\citep{hyp2}. 

\section{QMC and ChPT}

Recently, the concept of EFT is widely accepted. The scenario of low-energy QCD is 
characterized by pions and nucleons interacting through a force constrained by approximate chiral symmetry. 
In ChPT, the effective chiral Lagrangian relevant to the $N$-$N$ force is given by 
${\cal L}_{eff}={\cal L}^{(2)}_{\pi \pi} + {\cal L}^{(1)}_{\pi N} + {\cal L}^{(2)}_{\pi N} + \cdots$, where the superscript 
refers to the number of chiral dimension and the ellipsis stands for terms of chiral third order or higher~\citep{mach}.   
At the lowest and leading order, the $\pi \pi$ Lagrangian density is given by 
\bge
{\cal L}^{(2)}_{\pi \pi} = \frac{f_\pi^2}{4} \mbox{tr} \left[ \partial^\mu U \partial_\mu U^\dagger + m_\pi^2 (U + U^\dagger)  \right] , 
\label{pi-pi}
\ene
with $U= \exp[ i{\vec \tau} \cdot {\vec \phi}/f_\pi]$, and the $\pi N$ Lagrangian density is 
\bge
{\cal L}^{(1)}_{\pi N} = {\bar N} \left[ i \ \slash\!\!\!\!\!\!{\cal D} - M_N + 
\frac{g_A}{2} \gamma_\mu \gamma_5 {\vec \tau} \cdot (D^\mu {\vec \phi}) \right] N , 
\label{pi-N}
\ene
with the axial vector coupling constant, $g_A$.  Low energy constants, $c_i$, appear in 
the terms of dimension 2 or higher in ${\cal L}_{\pi N}$. 

In contrast, the nuclear force in the CQMC is described by the exchange of the $\sigma$ and $\omega$ mesons, which are 
introduced {\it by hand}.  However, even in the CQMC, instead of assuming the $\sigma$, 
it may be possible to obtain the intermediate attractive 
force by calculating the two-pion exchange (TPE) diagrams between two nucleons. 
In fact, the Lagrangian density, Eq.(\ref{chiralv-lag}), 
generates (or generalizes) the Weinberg Lagrangian~\citep{tony}. 

In ChPT, the attractive, central force, $V_c$, is calculated by the TPE contributions at third order.  In 
configuration space, it may be written by the dispersion relation~\citep{disp}
\bge
V_c(r) = \frac{1}{2\pi^2} \int_{2m_\pi}^\infty d\mu \frac{e^{-r\mu}}{r} \mu \eta_c(\mu) , 
\label{disp}
\ene
where $\eta_c(\mu)$ is the mass spectrum given by the imaginary part of the potential in momentum space, 
and it consists of terms with $g_A^4$ 
or $g_A^2 \times c_i$.  Thus, when a nucleon is located at ${\vec r}\,'$, the scalar field at ${\vec r}$ is given by 
\bge
\sigma({\vec r}, {\vec r}\,') = \frac{1}{(2\pi^2)^2} \int_{2m_\pi}^\infty d\mu \mu {\tilde \eta}_c(\mu) \int d{\vec k} \,
\frac{e^{i{\vec k}\cdot ({\vec r} - {\vec r}\,')}}{{\vec k}^2 + \mu^2} , 
\label{scalar-field}
\ene
where ${\tilde \eta}_c = \eta_c/g_A^2$.  We then find the scalar mean-field, ${\bar \sigma}$, created by 
a uniform distribution of nucleons with density $\rho_B$ as
\bge
{\bar \sigma} = \rho_B \int d{\vec r}\,' \sigma({\vec r}, {\vec r}\,') 
= \frac{2}{\pi} \rho_B \int_{2m_\pi}^\infty d\mu \frac{{\tilde \eta}_c(\mu)}{\mu}  .  
\label{mean-scalar-field}
\ene
When the Fermi motion of nucleons is considered, the result is given by Eq.(\ref{mean-scalar-field}) with 
the nucleon scalar density $\rho_s$, instead of $\rho_B$, in matter. 
As in the QMC, this constant mean-field acts on the quark inside a nucleon as an attractive force, and modifies the quark wavefunction.  
Such change generates a variation of $g_A^2$ depending on ${\bar \sigma}$.  Thus, we find the 
self-consistency condition for ${\bar \sigma}$ 
\bge
{\bar \sigma} 
= \frac{2}{\pi} \rho_s \int_{2m_\pi}^\infty d\mu \frac{{\tilde \eta}_c(\mu, g_A({\bar \sigma}))}{\mu}  
\propto - \rho_s \frac{g_A^2({\bar \sigma})}{m_\sigma^2} , 
\label{scc}
\ene
where, in the right hand side of Eq.(\ref{scc}), we assumed the simplest case, 
namely ${\tilde \eta}_c \propto - g_A^2({\bar \sigma}) \delta(\mu^2 - m_\sigma^2)$.  
In this case, the scalar polarizability is given by $g_A^2({\bar \sigma})$. 
Because of the decrease of $g_A$ in matter, the mean-central potential, 
${\bar V_c} (=g_A^2{\bar \sigma})$, is weakened for high $\rho_B$.  
In fact, the CQMC predicts 
$g_A(\rho_B)/g_A(0) \simeq 1 - b (\rho_B/\rho_0)$ with $b \simeq 0.14$~\citep{cqmc}. 
This effect is truly caused by the internal structure of nucleon, 
and concerns the saturation of nuclear matter. 
The reduction of the central force in matter may be favorable, because 
recent calculations for a nuclear matter show that the N$^3$LO potential based on EFT produces a deep overbinding 
at large $\rho_B$~\citep{over}. 

\section{Summary}

We have proposed a quark-meson coupling model with chiral symmetry. 
We have studied the effect of pion and gluon exchanges on the baryon mass in a nuclear matter, and also discussed  
the relationship between the CQMC and ChPT. 
The mean-central potential is modified by the internal structure change of nucleon. 
It may be possible to extend the CQMC to include $\Delta$-particles explicitly~\citep{cbm,tony}, and construct 
a $\Delta$-full EFT~\citep{mach} based on the quark model.


\begin{theacknowledgments}
I would like to thank Tony Thomas very much for the good collaborations from 1990. 
I also thank Wally Melnitchouk for, in spite of not being able to attend the workshop, allowing me to contribute to the proceedings. 
This work was supported by Academic Frontier Project (Holcs, Tokyo University of Science, 2008-2010) of MEXT. 
\end{theacknowledgments}

\bibliographystyle{aipproc}

\end{document}